\documentclass[journal]{IEEEtran}
\usepackage{cite, bm,amsmath,graphicx,amssymb,mathtools,amsthm,subcaption,footmisc,framed,booktabs,dsfont}
\usepackage{mathrsfs}
\usepackage{algorithm}
\usepackage{algpseudocode}
\usepackage[margin=0.8in]{geometry}
\usepackage{fancyhdr,bm}
\usepackage{threeparttable}
   \usepackage[english]{babel}
    \usepackage{url}
    \usepackage{tikz}
    \usepackage{bm}
\begin{document}
\newcommand{\thd}[1]{$#1$-th}
\newcommand{\rbold}{\mathbf{r}}
\newcommand{\pbold}{\mathbf{p}}
\newcommand{\Fbold}{\mathbf{F}}
\newcommand{\fbold}{\mathbf{f}}
\newcommand{\dt}[1]{d_{\bm{t};{{#1}}}}
\newcommand{\du}[1]{d_{{{#1}};\bm{r}}}
\newcommand{\wn}{\frac{2\pi}{\lambda}}
\newcommand{\EXP}{\mathbb{E}}
\newcommand{\ch}{\mathbf{h}}
\newcommand{\automath}[1]{\ifmmode#1\else$#1$}
\newcommand{\Cq}{ C_\theta }
\newcommand{\Cr}{C_r }
\newcommand{\diag}{\operatorname{diag}}
\newcommand{\deltaT}{\Delta \hat{\theta}}
\newcommand{\epsT}{\varepsilon_\theta}
\newcommand{\epsL}{\varepsilon_r}
\newcommand{\mepsT}{\epsilon_\theta}
\newcommand{\mepsL}{\epsilon_r}
\newcommand{\Msub}{M_{\rm sub}}
\newcommand{\cod}{\Delta C(\hatT,\theta,\hatD,r)}
\newcommand{\Capp}{{{C}}}
\newcommand{\cae}{{\mathcal{E}}}
\newcommand{\bfa}{{\mathbf{a}}}
\newcommand{\codew}{{\boldsymbol{w}}}
\newcommand{\rrg}{r_{\rm rg}}
\newcommand{\lrg}{\ell_{\rm rg}}
\newcommand{\trg}{\theta_{\rm rg}}
\newcommand{\rrgq}{\rho_{\rm rg}}
\newcommand{\cFullFun}{C(\epsL,\epsT)}
\newcommand{\frf}{\mathbf{F}_{\mathrm{RF}}}
\newcommand{\fbb}{\mathbf{F}_{\mathrm{BB}}}
\newcommand{\fbbk}[1]{\mathbf{f}_{\mathrm{BB},#1}}
\newcommand{\frfk}[1]{\mathbf{f}_{\mathrm{RF},#1}}
\newcommand{\argmin}{\operatorname{argmin}}

\newtheoremstyle{custom_style} 
  {\topsep} 
  {\topsep} 
  {\normalfont} 
  {} 
  {\bfseries} 
  {} 
  {.2em} 
  {\thmname{#1} {\normalfont\textbf{\thmnumber{#2}}} \thmnote{\normalfont(#3)}} 

  \newtheoremstyle{custom_style} 
  {\topsep} 
  {\topsep} 
  {\normalfont} 
  {} 
  {\bfseries} 
  {} 
  {.2em} 
  {\thmname{#1} {\normalfont\textbf{\thmnumber{#2}}} \thmnote{\normalfont(#3)\textbf{.}}} 

\theoremstyle{custom_style}

\newtheorem{theorem}{\emph{\underline{Theorem}}}
\newtheorem{acknowledgement}[theorem]{Acknowledgement}
\renewcommand{\algorithmicensure}{ \textbf{repeat:}}
\newtheorem{axiom}[theorem]{Axiom}
\newtheorem{case}[theorem]{Case}
\newtheorem{claim}[theorem]{Claim}
\newtheorem{conclusion}[theorem]{Conclusion}
\newtheorem{condition}[theorem]{Condition}
\newtheorem{conjecture}[theorem]{\emph{\underline{Conjecture}}}
\newtheorem{criterion}[theorem]{Criterion}
\newtheorem{definition}{\emph{\underline{Definition}}}
\newtheorem{exercise}[theorem]{Exercise}
\newtheorem{lemma}{\emph{\underline{Lemma}}}
\newtheorem{example}{\emph{\underline{Example}}}
\newtheorem{observation}{\emph{\underline{Observation}}}
\newtheorem{corollary}{\emph{\underline{Corollary}}}
\newtheorem{notation}[theorem]{Notation}
\newtheorem{problem}[theorem]{Problem}
\newtheorem{proposition}[theorem]{\emph{\underline{Proposition}}}
\newtheorem{solution}[theorem]{Solution}
\newtheorem{method}[theorem]{Method}
\newtheorem{summary}[theorem]{Summary}
\newtheorem{assumption}{\emph{{Assumption}}}
\newtheorem{remark}{\bf \emph{\underline{Remark}}}

\def\qed{$\Box$}
\def\QED{\mbox{\phantom{m}}\nolinebreak\hfill$\,\Box$}
\def
\endproof{\hspace*{\fill}~\qed
\par
\endtrivlist\unskip}
\def\endproof{\hspace*{\fill}~\qed\par\endtrivlist\vskip3pt}

\def\E{\mathsf{E}}
\def\eps{\varepsilon}
\def\Lsp{{\boldsymbol L}}
\def\Bsp{{\boldsymbol B}}
\def\lsp{{\boldsymbol\ell}}
\def\Ltsp{{\Lsp^2}}
\def\Lpsp{{\Lsp^p}}
\def\Linsp{{\Lsp^{\infty}}}
\def\LtR{{\Lsp^2(\Rst)}}
\def\ltZ{{\lsp^2(\Zst)}}
\def\ltsp{{\lsp^2}}
\def\ltZt{{\lsp^2(\Zst^{2})}}
\def\ninN{{n{\in}\Nst}}
\def\oh{{\frac{1}{2}}}
\def\grass{{\cal G}}
\def\ord{{\cal O}}
\def\dist{{d_G}}
\def\conj#1{{\overline#1}}
\def\ntoinf{{n \rightarrow \infty }}
\def\toinf{{\rightarrow \infty }}
\def\tozero{{\rightarrow 0 }}
\def\trace{{\operatorname{trace}}}
\def\ord{{\cal O}}
\def\UU{{\cal U}}
\def\rank{{\operatorname{rank}}}
\def\acos{{\operatorname{acos}}}

\def\SINR{\mathsf{SINR}}
\def\SNR{\mathsf{SNR}}
\def\SIR{\mathsf{SIR}}
\def\tSIR{\widetilde{\mathsf{SIR}}}
\def\Ei{\mathsf{Ei}}
\def\l{\left}
\def\r{\right}
\def\({\left(}
\def\){\right)}
\def\llargebra{\left\{}
\def\rlargebra{\right\}}
\def\lmidbra{\left[}
\def\rmidbra{\right]}
\def\labs{\left|}
\def\rabs{\right|}

\setcounter{page}{1}

\newcommand{\eref}[1]{(\ref{#1})}
\newcommand{\fig}[1]{Fig.\ \ref{#1}}

\def\bydef{:=}
\def\ba{{\mathbf{a}}}
\def\bb{{\mathbf{b}}}
\def\bc{{\mathbf{c}}}
\def\bd{{\mathbf{d}}}
\def\bee{{\mathbf{e}}}
\def\bff{{\mathbf{f}}}
\def\bg{{\mathbf{g}}}
\def\bh{{\mathbf{h}}}
\def\bi{{\mathbf{i}}}
\def\bj{{\mathbf{j}}}
\def\bk{{\mathbf{k}}}
\def\bl{{\mathbf{l}}}
\def\bn{{\mathbf{n}}}
\def\bo{{\mathbf{o}}}
\def\bp{{\mathbf{p}}}
\def\bq{{\mathbf{q}}}
\def\br{{\mathbf{r}}}
\def\bs{{\mathbf{s}}}
\def\bt{{\mathbf{t}}}
\def\bu{{\mathbf{u}}}
\def\bv{{\mathbf{v}}}
\def\bw{{\mathbf{w}}}
\def\bx{{\mathbf{x}}}
\def\by{{\mathbf{y}}}
\def\bz{{\mathbf{z}}}
\def\b0{{\mathbf{0}}}

\def\bA{{\mathbf{A}}}
\def\bB{{\mathbf{B}}}
\def\bC{{\mathbf{C}}}
\def\bD{{\mathbf{D}}}
\def\bE{{\mathbf{E}}}
\def\bF{{\mathbf{F}}}
\def\bG{{\mathbf{G}}}
\def\bH{{\mathbf{H}}}
\def\bI{{\mathbf{I}}}
\def\bJ{{\mathbf{J}}}
\def\bK{{\mathbf{K}}}
\def\bL{{\mathbf{L}}}
\def\bM{{\mathbf{M}}}
\def\bN{{\mathbf{N}}}
\def\bO{{\mathbf{O}}}
\def\bP{{\mathbf{P}}}
\def\bQ{{\mathbf{Q}}}
\def\bR{{\mathbf{R}}}
\def\bS{{\mathbf{S}}}
\def\bT{{\mathbf{T}}}
\def\bU{{\mathbf{U}}}
\def\bV{{\mathbf{V}}}
\def\bW{{\mathbf{W}}}
\def\bX{{\mathbf{X}}}
\def\bY{{\mathbf{Y}}}
\def\bZ{{\mathbf{Z}}}

\def\mA{{\mathbb{A}}}
\def\mB{{\mathbb{B}}}
\def\mC{{\mathbb{C}}}
\def\mD{{\mathbb{D}}}
\def\mE{{\mathbb{E}}}
\def\mF{{\mathbb{F}}}
\def\mG{{\mathbb{G}}}
\def\mH{{\mathbb{H}}}
\def\mI{{\mathbb{I}}}
\def\mJ{{\mathbb{J}}}
\def\mK{{\mathbb{K}}}
\def\mL{{\mathbb{L}}}
\def\mM{{\mathbb{M}}}
\def\mN{{\mathbb{N}}}
\def\mO{{\mathbb{O}}}
\def\mP{{\mathbb{P}}}
\def\mQ{{\mathbb{Q}}}
\def\mR{{\mathbb{R}}}
\def\mS{{\mathbb{S}}}
\def\mT{{\mathbb{T}}}
\def\mU{{\mathbb{U}}}
\def\mV{{\mathbb{V}}}
\def\mW{{\mathbb{W}}}
\def\mX{{\mathbb{X}}}
\def\mY{{\mathbb{Y}}}
\def\mZ{{\mathbb{Z}}}

\def\cA{\mathcal{A}}
\def\cB{\mathcal{B}}
\def\cC{\mathcal{C}}
\def\cD{\mathcal{D}}
\def\cE{\mathcal{E}}
\def\cF{\mathcal{F}}
\def\cG{\mathcal{G}}
\def\cH{\mathcal{H}}
\def\cI{\mathcal{I}}
\def\cJ{\mathcal{J}}
\def\cK{\mathcal{K}}
\def\cL{\mathcal{L}}
\def\cM{\mathcal{M}}
\def\cN{\mathcal{N}}
\def\cO{\mathcal{O}}
\def\cP{\mathcal{P}}
\def\cQ{\mathcal{Q}}
\def\cR{\mathcal{R}}
\def\cS{\mathcal{S}}
\def\cT{\mathcal{T}}
\def\cU{\mathcal{U}}
\def\cV{\mathcal{V}}
\def\cW{\mathcal{W}}
\def\cX{\mathcal{X}}
\def\cY{\mathcal{Y}}
\def\cZ{\mathcal{Z}}
\def\cd{\mathcal{d}}
\def\Mt{M_{t}}
\def\Mr{M_{r}}
\def\O{\Omega_{M_{t}}}
\newcommand{\figref}[1]{{Fig.}~\ref{#1}}
\newcommand{\tabref}[1]{{Table}~\ref{#1}}

\newcommand{\var}{\mathsf{var}}
\newcommand{\fb}{\tx{fb}}
\newcommand{\nf}{\tx{nf}}
\newcommand{\BC}{\tx{(bc)}}
\newcommand{\MAC}{\tx{(mac)}}
\newcommand{\Pout}{p_{\mathsf{out}}}
\newcommand{\nnn}{\nn\\}
\newcommand{\FB}{\tx{FB}}
\newcommand{\TX}{\tx{TX}}
\newcommand{\RX}{\tx{RX}}
\renewcommand{\mod}{\tx{mod}}
\newcommand{\m}[1]{\mathbf{#1}}
\newcommand{\td}[1]{\tilde{#1}}
\newcommand{\sbf}[1]{\scriptsize{\textbf{#1}}}
\newcommand{\stxt}[1]{\scriptsize{\textrm{#1}}}
\newcommand{\suml}[2]{\sum\limits_{#1}^{#2}}
\newcommand{\sumlk}{\sum\limits_{k=0}^{K-1}}
\newcommand{\eqhsp}{\hspace{10 pt}}
\newcommand{\tx}[1]{\texttt{#1}}
\newcommand{\Hz}{\ \tx{Hz}}
\newcommand{\sinc}{\tx{sinc}}
\newcommand{\tr}{\mathrm{tr}}
\newcommand{\MAI}{\tx{MAI}}
\newcommand{\ISI}{\tx{ISI}}
\newcommand{\IBI}{\tx{IBI}}
\newcommand{\CN}{\tx{CN}}
\newcommand{\CP}{\tx{CP}}
\newcommand{\ZP}{\tx{ZP}}
\newcommand{\ZF}{\tx{ZF}}
\newcommand{\SP}{\tx{SP}}
\newcommand{\MMSE}{\tx{MMSE}}
\newcommand{\MINF}{\tx{MINF}}
\newcommand{\RC}{\tx{MP}}
\newcommand{\MBER}{\tx{MBER}}
\newcommand{\MSNR}{\tx{MSNR}}
\newcommand{\MCAP}{\tx{MCAP}}
\newcommand{\vol}{\tx{vol}}
\newcommand{\ah}{\hat{g}}
\newcommand{\tg}{\tilde{g}}
\newcommand{\teta}{\tilde{\eta}}
\newcommand{\heta}{\hat{\eta}}
\newcommand{\uh}{\m{\hat{s}}}
\newcommand{\eh}{\m{\hat{\eta}}}
\newcommand{\hv}{\m{h}}
\newcommand{\hh}{\m{\hat{h}}}
\newcommand{\Po}{P_{\mathrm{out}}}
\newcommand{\Poh}{\hat{P}_{\mathrm{out}}}
\newcommand{\Ph}{\hat{\gamma}}
\newcommand{\mat}[1]{\begin{matrix}#1\end{matrix}}
\newcommand{\ud}{^{\dagger}}
\newcommand{\C}{\mathcal{C}}
\newcommand{\nn}{\nonumber}
\newcommand{\nInf}{U\rightarrow \infty}

\title{Position Optimization for Two-layer Movable Antenna Systems}

\author{Liujia~Yao,~\IEEEmembership{Student~Member,~IEEE}, Changsheng~You,~\IEEEmembership{Member,~IEEE}, Chao~Zhou,~\IEEEmembership{Student~Member,~IEEE}, Beixiong~Zheng,~\IEEEmembership{Senior Member,~IEEE}, and Weidong~Mei,~\IEEEmembership{Member,~IEEE}%
\thanks{L. Yao, C. You, and C. Zhou are with the Department of Electrical and Electronic Engineering, Southern University of Science and Technology, Shenzhen, China (e-mail: yaolj2024@mail.sustech.edu.cn; zhouchao2024@mail.sustech.edu.cn; youcs@sustech.edu.cn).
\\
\indent \indent B. Zheng is with the School of Microelectronics, South China University of Technology, Guangzhou, China (e-mail: bxzheng@scut.edu.cn). \\
\indent \indent W. Mei is with the National Key Laboratory of Wireless Communications, University of Electronic Science and Technology of China, Chengdu, China (e-mail: wmei@uestc.edu.cn). \\
\indent \indent \textit{(Corresponding author: Changsheng You)}}%
\vspace{-2.8em}
}

\maketitle 

\begin{abstract}
Movable antenna (MA) is a promising technology for improving the performance of wireless communication systems by providing new degrees-of-freedom (DoFs) in antenna position optimization.
However, existing works on MA systems have mostly considered element-wise \emph{single-layer} MA (SL-MA) arrays, where all the MAs move within the given movable region, hence inevitably incurring high control complexity and hardware cost in practice.
To address this issue, we propose in this letter a new \emph{two-layer} MA array (TL-MA), where the positions of MAs are jointly determined by the \emph{large-scale movement} of multiple subarrays and the \emph{small-scale fine-tuning} of per-subarray MAs. 
In particular, an optimization problem is formulated to maximize the sum-rate of the TL-MA-aided communication system by jointly optimizing the subarray-positions, per-subarray (relative) MA positions, and receive beamforming.  
To solve this non-convex problem, we propose an alternating optimization (AO)-based particle swarm optimization (PSO) algorithm, which alternately optimizes the positions of subarrays and per-subarray MAs, given the optimal receive beamforming. 
Numerical results verify that the proposed TL-MA significantly reduces the \textit{sum-displacement} of MA motors (i.e., the total moving distances of all motors) of element-wise SL-MA, while achieving comparable rate performance.
\end{abstract}
\vspace{-0.2em}
\vspace{-1em}
\section{Introduction}
With the wide deployment of the fifth-generation (5G) mobile communication networks, the next-generation wireless networks are expected to deliver orders-of-magnitude improvement in terms of network capacity, reliability, and latency reduction~\cite{you2024generationadvancedtransceivertechnologies}.
To this end, a variety of new antenna technologies have been recently proposed, e.g., fluid antenna systems (FAS)~\cite{FAStotorial}, (6D) movable antennas (MA)~\cite{graph,RA}, rotatable antennas (RA)~\cite{rotatableAntennaModelingOptimization2025}, and pinching antenna systems (PASS)~\cite{pinchingA}. Unlike the existing multiple-input multiple-output (MIMO) systems based on fixed-position antennas (FPA), these position/rotation-agile antenna technologies provide an additional spatial degrees-of-freedom (DoFs) by dynamically adjusting the positions/rotations of antennas within a predefined region, thereby achieving better rate performance than conventional FPA systems~\cite{totarial}. 

\begin{figure}
    \centering
    \begin{subfigure}[b]{0.9\linewidth}
        \centering
        \includegraphics[width=\linewidth]{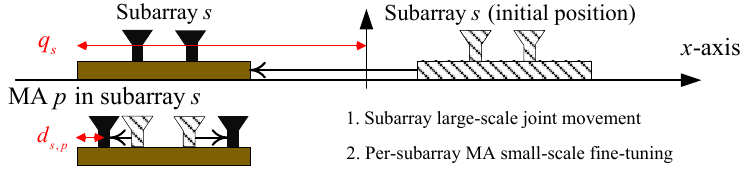}
        \caption{The \textit{two-layer} antenna movement of TL-MA.}
        \label{fig:system_tlma}
    \end{subfigure}
    \vspace{-4pt}
    \begin{subfigure}[b]{0.9\linewidth}
        \centering
        \includegraphics[width=\linewidth]{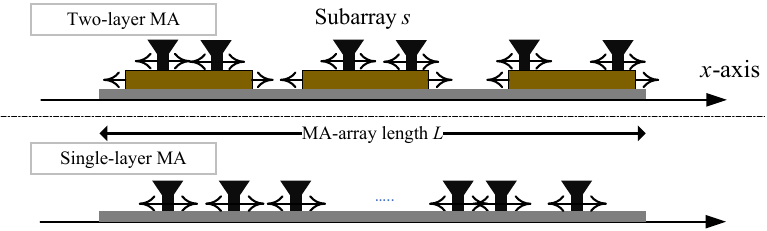}
        \caption{Comparison between SL-MAs and the proposed TL-MA.}
        \label{fig:system_compare}
    \end{subfigure}
    \caption{\small Illustration of the proposed TL-MA and its comparison with conventional SL-MA. }\vspace{-2em}
    \label{fig:system}
\end{figure}

Among others, MA technology has attracted growing research attention owing to its capability to provide flexible antenna position adjustment adaptive to dynamic environments. For example, the authors in~\cite{maMIMOCapacityCharacterization2024} characterized the capacity of MA-aided MIMO systems for the single-user scenario, which was then extended  to the multi-user MIMO scenario in~\cite{piMultiuserCommunicationsMovableAntenna2023a}. 
While the performance gains of MAs have been verified in~\cite{maMIMOCapacityCharacterization2024,piMultiuserCommunicationsMovableAntenna2023a}, most existing works considered element-wise single-layer MA (SL-MA) architectures, as illustrated in Fig.~\ref{fig:system}(\subref{fig:system_compare}). Such designs, however, may incur high hardware cost and implementation complexity in practice, since each MA element needs to be attached with a high-speed actuator (e.g., 10–20 m/s~\cite{MAimplementation}) and their position optimization requires sophisticated path planning accounting for collision avoidance~\cite{MAimplementation}.

To achieve a more favorable trade-off between rate performance and hardware complexity, we propose in this letter a new \emph{two-layer} MA (TL-MA) architecture as shown in Fig.~\ref{fig:system}(b). 
Specifically, this architecture employs a hierarchical positioning adjustment mechanism, where the positions of all MAs are jointly determined by the \emph{large-scale movement} of subarrays as well as \emph{small-scale fine-tuning} of per-subarray MAs.
Specifically, we formulate an optimization problem to maximize the system sum-rate by jointly designing the positions of subarrays and MAs given the optimal receive beamforming. This high-dimensional non-convex problem is then solved by an efficient alternating optimization (AO)-based particle swarm optimization (PSO) algorithm.
Numerical results show that the proposed TL-MA achieves comparable rate performance with SL-MA, yet requiring significantly smaller sum-displacement (and thus lower complexity).
Moreover, the proposed AO-based PSO algorithm achieves better rate performance over the benchmark under the all-at-once antenna position optimization. 
In addition, it is worth noting that, compared to the existing array-wise MA architecture that adjusts the positions of subarrays only~\cite{gma}, our proposed TL-MA architecture also incorporates small-scale fine-tuning of per-subarray antennas, thus achieving better rate performance.
\vspace{-1em}
\section{System model and problem formulation}
\label{sec:system_model}
We consider a narrow-band multi-user MIMO system, where a new TL-MA is employed at the base station (BS) to serve $K$ single-antenna users in the uplink under a quasi-static fading channel. 
\vspace{-1.5em}
\subsection{System Model}
\textbf{\underline{TL-MA Model:}} The proposed TL-MA array shown in Fig.~\ref{fig:system}(b) consists of $M=M_{\rm S}\times M_{\rm A}$ MAs in total, which are grouped into $M_{\rm S}$ movable subarrays (denoted as $\mathcal{M}_{\rm S}\triangleq \{1,...,M_{\rm S}\}$), each comprising ${M}_{\rm A}$ MAs (denoted as $\mathcal{M}_{\rm A}\triangleq\{1,...,M_{\rm A}\}$). 
As such, the positions of $M$ MAs are jointly determined by the \emph{large-scale} movement of subarrays and the \emph{small-scale} fine-tuning of per-subarray MAs, as shown in Fig.~\ref{fig:system}(a). Specifically, the $M_{\rm S}$ subarrays have the same array length of $L_{\rm A}$ and can move flexibly in a prescribed movable region of length $L$ (named as MA-array length, see Fig.~\ref{fig:system}(\subref{fig:system_compare})). 
The position of the $s$-th subarray, with $s\in \mathcal{M}_{\rm S}$, can be determined by parameters $\{L_{\rm A}, q_s\}$, where $q_s$ denotes the $x$-axis coordinate of the starting point of the subarray $s$. 
Moreover, for subarray $s$, its MAs can adjust their positions in the interval $[q_s,q_s+L_{\rm A}]$ at the $x$-axis. 
In addition, for the $a$-th ($a\in \mathcal{M}_{\rm A}$) MA of subarray $s$, we denote by $d_{s,a}$ its \emph{relative} distance from $q_{s}$ (i.e., the starting point of its subarray). Then, its (absolute) position on the $x$-axis, denoted by $\delta_{s,a}$, is given by $\delta_{s,a}=q_{s}+d_{s,a}$. 
By defining $\mathbf{d}_s\triangleq \{d_{s,a}\}_{a=1}^{M_{\rm A}} \in \mathbb{R}^{M_{\rm A} \times 1}$, $\mathbf{q} \triangleq \{q_s\}_{s=1}^{M_{\rm S}}\in \mathbb{R}^{M_{\rm S}\times 1}$, $\bm{\delta}_s\triangleq [\delta_{s,1},...,\delta_{s,M_{\rm A}}]^T \in \mathbb{R}^{M_{\rm A}\times 1}$, and
 $\bm{\delta} \triangleq [\bm{\delta}_1^T,...,\bm{\delta}_{M_{\rm S}}^T]^T \in \mathbb{R}^{M\times 1}$, we obtain 
 \begin{equation}
    \bm{\delta}(\mathbf{q},\{\mathbf{d}_s\}) = \bigl(\mathbf{q} \otimes \mathbf{1}_{M_{\rm A}}\bigr) + [\mathbf{d}_1^T,...,\mathbf{d}_{M_{\rm S}}^T]^T.
 \end{equation}

Unlike SL-MA which directly moves all MAs, TL-MA first moves subarrays (i.e., $\mathbf{q}$) and then fine-tunes MAs (i.e., $\{\mathbf{d}_s\}$), reducing implementation complexity (see Fig.~\ref{fig:system}).
For the TL-MA movement, the following constraints need to be satisfied:  First, to prevent the subarrays from overlapping with each other or exceeding the movable regions, we have
\begin{subequations}
    \begin{align}
 \label{eq:cons_inter_array}
 & q_{s+1} - q_s \ge L_{\rm A}, \quad\; \forall s=1,2,\dots,M_{\rm S}-1, \\
 \label{eq:cons_boundary_array}
 -&\frac{L}{2} \le q_s \le \frac{L}{2}-L_{\rm A}, \forall s=1,2,\dots,M_{\rm S}.
    \end{align}
\end{subequations}
On the other hand, for each subarray $s$, the relative MA positions are constrained to prevent MAs from moving out of their subarray region and to ensure a minimum half-wavelength spacing, which are given by
\begin{subequations}
    \begin{align}
 \label{eq:cons_intra_antenna}
 & \frac{\lambda}{4} \le d_{s,a} \le L_{\rm A} - \frac{\lambda}{4}, \forall a=1,2,\dots, M_{\rm A}, \forall s,\\
 \label{eq:cons_intrq}
 & d_{s,a+1} - d_{s,a} \ge \frac{\lambda}{2}, \;\;\; \forall a=1,2,\dots,M_{\rm A}-1, \forall s.
    \end{align}
\end{subequations}

\textbf{\underline{Channel Model:}} Let $\ch_k(\mathbf{q},\{\mathbf{d}_s\})\in \mathbb{C}^{M\times 1}, \forall k \in \mathcal{K}$ denote the channel from user $k$ to the TL-MA array with $\mathcal{K} \triangleq \{1, \ldots, K\}$, which is assumed to be perfectly known\footnote{
    The existing MA channel estimation methods (see, e.g.,~\cite{totarial}) can be applied to acquire the channel state information (CSI).
    For the case with imperfect CSI, robust beamforming methods can be applied to guarantee the worst-case performance~\cite{robust}, which is left for our future work.
    } at the BS. Specifically, the channel $\mathbf{h}_k$ is modeled as
\begin{equation}
 \label{eq:channel_model}
    \ch_k^H (\mathbf{q},\{\mathbf{d}_s\}) = \sum_{n=1}^{N_{\rm PA}} \beta_{k,n}  \mathbf{b}^H(\theta_{k,n},\mathbf{q},\{\mathbf{d}_s\}),
\end{equation}
where $\beta_{k,n}$ is the complex-valued channel gain of the $n$-th path for user $k$, $N_{\rm PA}$ is the number of paths, and $\theta_{k,n}$ is the spatial angle between the $n$-th scatterer of user $k$ and the MA-array. Herein, $\mathbf{b}(\theta_{k,n},\mathbf{q},\{\mathbf{d}_s\})\in \mathbb{C}^{M\times 1}$ denotes the far-field channel steering vector, which is given by
\begin{equation}
 \label{eq:steering_vector}
    \mathbf{b}(\theta_{k,n},\mathbf{q},\{\mathbf{d}_s\}) = [e^{-\jmath \frac{2\pi}{\lambda}\delta_1 \theta_{k,n}}, \ldots, e^{-\jmath \frac{2\pi}{\lambda}\delta_M \theta_{k,n}}]^T,
\end{equation}
where $\lambda$ is the carrier wavelength and $\delta_m\triangleq [\bm{\delta}]_m$. 

\textbf{\underline{Signal Model:}} Let $x_k$ denote the data symbol transmitted by user $k$, which satisfies $\mathbb{E}[x_k^H x_k] = P_{\rm t},\forall k\in \mathcal{K}$, with $P_{\rm t}$ representing its transmit power. As such, the received signal at the BS can be expressed as
\begin{equation}
    {\mathbf{y}}= \mathbf{H}\mathbf{x} + \mathbf{z},
\end{equation} 
where $\mathbf{H} = \big[\ch_1(\mathbf{q},\{\mathbf{d}_s\}),...,\ch_K(\mathbf{q},\{\mathbf{d}_s\})\big] \in \mathbb{C}^{M\times K}$ is the channel matrix, $\mathbf{x} = [x_1,...,x_K]^T \in \mathbb{C}^{K \times 1}$ is the transmit signal vector, and $\mathbf{z} \sim \mathcal{CN}(0,\sigma^2 \mathbf{I})$ is the received additive white Gaussian noise (AWGN) vector at the BS. To decode the uplink signal of each user $k$, the receive beamforming vector $\mathbf{v}_k\in \mathbb{C}^{M\times 1}$ is applied at the BS with $\|\mathbf{v}_k\|^2=1$. As such, the decoded signal from the $k$-th user is given by 
\begin{equation}
    \hat{x}_k = \mathbf{v}_k^H {\mathbf{y}} = \mathbf{v}_k ^H \mathbf{h}_k x_k + \mathbf{v}_k^H \sum_{ i\ne k}^{K} \mathbf{h}_i x_i + \mathbf{v}_k^H \mathbf{z}, \forall k\in \mathcal{K},
\end{equation} 
and its achievable rate in bits/second/Hertz (bps/Hz) is
\begin{equation}
    \begin{aligned}
 \label{eq:rate}
         \hspace{-8pt}R_k(\mathbf{q},\{\mathbf{d}_s\},\{\mathbf{v}_k\})\! =\! \log_2\left(1\!+\!\frac{\gamma |\mathbf{v}_k^H \mathbf{h}_k|^2 }{\gamma \sum_{i\ne k}^{K}|\mathbf{v}_k^H \mathbf{h}_i|^2 + 1}\right),\!\!\!
    \end{aligned}
    \vspace{-0.2em}
\end{equation}
where $\gamma = \frac{P_{\rm t}}{\sigma^2}$ is the transmit signal-to-noise ratio (SNR).

\textbf{\underline{Displacement model:}} 
To evaluate the implementation complexity of different MA hardware architectures, we consider a new metric called \emph{sum-displacement}, which characterizes the total moving distances of all MA motors, and represents one of the main metrics of the system cost (e.g., hardware cost, movement control complexity, energy consumption)~\cite{gma,MAimplementation}.
Specifically, for the proposed TL-MA, its sum-displacement of all subarrays is $C_{\rm S} = \sum_{s} |q^{(0)}_s - q^{*}_s|$, where $q^{(0)}_s$ (or $q^{*}_s$) are the initial (or optimized) positions of the subarrays, respectively. In addition, the sum-displacement of all MAs in movement fine-tuning is $C_{\rm A} = \sum_{s} \sum_{a} |d^{(0)}_{s,a} - d^{*}_{s,a}|$, 
where $d^{(0)}_{s,a}$ (or $d^{*}_{s,a}$) are the initial (or optimized) relative positions of the MAs in subarray $s$, respectively. 

\vspace{-1.5em}
\subsection{Problem Formulation}
We aim to maximize the uplink sum-rate\footnote{
This work focuses on the uplink sum-rate maximization problem under the given TL-MA architecture (i.e., the number of subarrays and per-subarray length).
For the more general objective that balances sum-rate with the associated sum-displacement cost, the problem can be formulated as a multi-objective optimization problem. The Pareto-optimal solutions can be obtained via exhaustive search, the details of which are left for future work.
} by jointly optimizing the subarray-positions $\mathbf{q}$, the per-subarray \emph{relative} MA-positions $\{\mathbf{d}_s\}$, and the receive beamforming $\{\mathbf{v}_k\}$. This optimization problem can be formulated as
\begin{subequations}
    \begin{align}
 \label{eq:problem}
 (\text{P1})\quad &\max_{\mathbf{q},\{\mathbf{d}_s\},\{\mathbf{v}_k\}} \quad \sum_{k=1}^{K} R_k(\mathbf{q},\{\mathbf{d}_s\},\{\mathbf{v}_k\}) \\
 & \;\;\;\;\;\text{s.t.} \quad\quad \;\;\text{\eqref{eq:cons_inter_array}, \eqref{eq:cons_boundary_array}, \eqref{eq:cons_intra_antenna}, \eqref{eq:cons_intrq}}, \notag\\
 & \phantom{\;\;\;\text{s.t.}}\quad\;\;\quad \;\; \|\mathbf{v}_k\|^2 = 1, \forall k\in\mathcal{K}, \label{eq:cons_precoding} 
    \end{align}
\end{subequations}
where \eqref{eq:cons_precoding} is the unit-power constraint for receive beamforming. 
Problem (P1) is generally hard to solve optimally due to its non-concave objective function and non-convex constraints. 
Moreover, the coupled two-layer position optimization variables (i.e., $\mathbf{q}$ and $\{\mathbf{d}_s\}$) make problem (P1) even more challenging, rendering existing MA-position optimization methods, such as successive convex approximation (SCA)~\cite{maMIMOCapacityCharacterization2024} and PSO~\cite{piMultiuserCommunicationsMovableAntenna2023a}, much less effective.
\vspace{-1em}
\section{Proposed Solution to Problem (P1)}
In this section, we propose an efficient algorithm to solve the sum-rate maximization problem (P1). Specifically, given arbitrary subarray positions $\mathbf{q}$ and relative MA positions $\{ \mathbf{d}_{s} \}$, we first obtain the optimal receive beamforming $\{\mathbf{v}_{k}\}$ in closed form, based on which the original problem (P1) is reformulated as an equivalent problem for optimizing $\mathbf{q}$ and $\{ \mathbf{d}_{s} \}$. Then we further develop an AO-based algorithm to iteratively optimize these two-layer position variables until the convergence is achieved.
\vspace{-1em}
\subsection{Optimization of Receive Beamforming}
\label{sec:opt_precoding}
Given any subarray-positions $\mathbf{q}$ and relative MA-positions $\{\mathbf{d}_s\}$, problem (P1) reduces to optimizing the receive beamforming only for maximizing the signal-to-interference-plus-noise ratio (SINR) of each user $k$, which is given by
\begin{equation}
 \label{eq:SINR_rly}
    \SINR_k = \frac{\gamma |\mathbf{h}_k^H \mathbf{v}_k|^2 }{\gamma \sum_{i\ne k}^{K}|\mathbf{h}_i^H \mathbf{v}_k|^2 + 1}=\frac{\gamma \mathbf{v}_k^H \mathbf{h}_k \mathbf{h}_k^H \mathbf{v}_k}{\mathbf{v}_k^H \mathbf{C}_k \mathbf{v}_k},\forall k \in \mathcal{K},
\end{equation}
with $\mathbf{C}_k \triangleq \gamma \sum_{i\ne k}^{K} \mathbf{h}_i \mathbf{h}_i^H + \mathbf{I}$ being the interference-plus-noise covariance matrix for user $k$. 
By employing minimum mean square error (MMSE) receive beamforming~\cite{gma}, the maximum SINR for user $k$ can be obtained, whose MMSE receive beamforming is 
\begin{equation}
 \label{eq:problem_precoding_individual}
    \mathbf{v}_k^* = \operatorname*{argmax}_{\|\mathbf{v}_k\|^2=1} \frac{\gamma \mathbf{v}_k^H \mathbf{h}_k  \mathbf{h}_k^H \mathbf{v}_k }{\mathbf{v}_k^H \mathbf{C}_k \mathbf{v}_k} = \frac{\mathbf{C}_k^{-1} \mathbf{h}_k}{\|\mathbf{C}_k^{-1} \mathbf{h}_k\|}, \forall k\in\mathcal{K}.
\end{equation}
\vspace{-1em}

\subsection{Joint Position Optimization of Subarrays and Antennas}
By substituting the optimal receive beamforming $\{\mathbf{v}_k^*\}$ into \eqref{eq:problem_precoding_individual}, 
the original problem (P1) can be reformulated as the following problem for optimizing the subarray-positions $\mathbf{q}$ and \textit{relative} MA-positions $\{\mathbf{d}_s\}$ (named as MA-positions in the sequel for brevity)
\begin{subequations}
    \begin{align}
 \label{eq:problem_fixed_precoding}
 (\text{P2})\quad &\max_{\mathbf{q},\{\mathbf{d}_s\}} \quad \tilde{R}(\mathbf{q},\{\mathbf{d}_s\}) \\
 & \;\text{s.t.} \quad\;\; \;\;\text{\eqref{eq:cons_inter_array}, \eqref{eq:cons_boundary_array}, \eqref{eq:cons_intra_antenna}, \eqref{eq:cons_intrq}}\notag,
    \end{align}
\end{subequations}
where $\tilde{R}(\mathbf{q},\{\mathbf{d}_s\})\triangleq \sum_{k=1}^{K} R_k(\mathbf{q},\{\mathbf{d}_s\},\{\mathbf{v}_k^*\})$ is the sum-rate under the optimal receive beamforming.
However, problem~(P2) is still a non-convex problem and hard to solve optimally due to the non-concave objective function~\eqref{eq:problem_fixed_precoding}. 
The high dimensionality of the optimization variables $(\mathbf{q},\{\mathbf{d}_s\})$ also makes this problem even more challenging.
To tackle these difficulties, we propose an efficient AO-based PSO algorithm to solve problem (P2).
Specifically, it alternately optimizes the subarray-positions $\mathbf{q}$ and per-subarray MA-positions $\{\mathbf{d}_s\}$ with the other being fixed, until the convergence is achieved, whose details are presented below.
\subsubsection{Optimization of Subarray-Positions}
\label{sec:proposed_algo}
\label{sec:opt_sub_array}

Given any MA-positions $\{\mathbf{d}_s\}$, problem (P2) reduces to the following problem for optimizing the subarray-positions $\mathbf{q}$
\begin{equation}
 (\text{P3})\quad \max_{\mathbf{q}} \quad \tilde{R}(\mathbf{q},\{\mathbf{d}_s\}) 
  \;\;\;\text{s.t.} \;\;\text{\eqref{eq:cons_inter_array}, \eqref{eq:cons_boundary_array}}. \notag
\end{equation}
Problem (P3) is challenging to solve optimally because of the non-concave objective function $\tilde{R}(\mathbf{q},\{\mathbf{d}_s\})$. 
To address this issue, the PSO algorithm is employed to obtain a high-quality solution to problem (P3), with the details for updating subarray-positions $\mathbf{q}$ presented below~\cite{piMultiuserCommunicationsMovableAntenna2023a}. 

\textbf{Initialization:} At the beginning of the PSO algorithm, a swarm of $I_{\rm P}$ (whose index set is $\mathcal{I}_{\rm P}\triangleq \{1,2,...,I_{\rm P}\}$) feasible position vectors are generated. The position of particle $i$ is denoted by $\mathbf{q}^{(0)}_{i}\in \mathbb{R}^{M_{\rm S}\times 1}, i \in \mathcal{I}_{\rm P}$, with $\mathbf{s}_{i}^{(0)}\in \mathbb{R}^{M_{\rm S}\times 1}, i \in \mathcal{I}_{\rm P}$ denoting its initial speed, which is adjusted to search for high-quality subarray-positions.

\textbf{Fitness Function Design:} To evaluate the quality of any subarray-position $\mathbf{q}$, the following fitness function is considered~\cite{piMultiuserCommunicationsMovableAntenna2023a}
\begin{equation}
 \label{eq:fitness_function}
    \mathcal{F}_{\rm S}(\mathbf{q}) = \tilde{R}(\mathbf{q},\{\mathbf{d}_s\}) - \kappa P_{\rm S}(\mathbf{q}),
\end{equation}
where the first term corresponds to the objective function in problem (P3) and the second term is the subarray-wise penalty function accounting for constraints \eqref{eq:cons_inter_array} and \eqref{eq:cons_boundary_array}. Specifically,  $\kappa$ is the penalty coefficient and $P_{\rm S}(\mathbf{q})$ is
\begin{align}
 P_{\rm S}(\mathbf{q}) &= \sum_{s=1}^{M_{\rm S}-1}\left[L_{\rm A}-(q_{s+1}-q_s)\right]^+ \notag\\
 &+\sum_{s=1}^{M_{\rm S}} \left(\!\!\left[-\frac{L}{2}-q_s\right]^+ \!\!\!\!\!\! +\!\! \left[q_s - \left(\frac{L}{2}-L_{\rm A}\right)\right]^+\!\right),\label{eq:penalty_function}
\end{align}
where $[x]^+\triangleq \max(0,x)$.
Note that the first and second terms in~\eqref{eq:penalty_function} respectively characterize the extent of violation for constraints~\eqref{eq:cons_inter_array} and~\eqref{eq:cons_boundary_array} for subarray positions $\mathbf{q}$.


\textbf{Positions Update:} Let $t$ denote the PSO iteration index, with $\mathcal{T} \triangleq \{1, 2, ..., t\}$.
In each PSO iteration, the update of the position is determined by the personal-best position of particle $i$ (among its personal history) from the previous iteration, denoted by $\mathbf{q}_{i,{\rm pbest}}^{(t-1)}\in \mathbb{R}^{M_{\rm S}\times 1}$, and the global-best position (among the particle swarm) from the previous iteration, denoted by $\mathbf{q}_{\rm gbest}^{(t-1)}\in \mathbb{R}^{M_{\rm S}\times 1}$. 
Specifically, the personal/global-best positions are respectively given by
\begin{subequations}
    \begin{align}
    \label{eq:best_position_personal}
    &\mathbf{q}_{i,{\rm pbest}}^{(t-1)} = \operatorname*{argmax}_{\tau\in \mathcal{T}\setminus \{t\}} \mathcal{F}_{\rm S}(\mathbf{q}_i^{(\tau)}), \forall i \in \mathcal{I}_{\rm P},\\
    \label{eq:best_position_global}
    &\mathbf{q}_{\rm gbest}^{(t-1)} \;\,= \operatorname*{argmax}_{i \in \mathcal{I}_{\rm P}} \mathcal{F}_{\rm S}(\mathbf{q}_{i,{\rm pbest}}^{(t-1)}).
    \end{align}
\end{subequations}
Given the obtained personal-best and global-best positions, the speed of the $i$-th particle is updated as follows
\begin{align}
    \label{eq:speed_update}
    \mathbf{s}_i^{(t)} & = \omega \mathbf{s}_i^{(t-1)} 
    + c_1 \mathbf{r}_1^{(t-1)} \odot \big( \mathbf{q}_{i,{\rm pbest}}^{(t-1)} - \mathbf{q}_i^{(t-1)} \big) \notag \\
    & \quad + c_2 \mathbf{r}_2^{(t-1)} \odot \big( \mathbf{q}_{\rm gbest}^{(t-1)} - \mathbf{q}_i^{(t-1)} \big), \forall i \in \mathcal{I}_{\rm P},
\end{align}
where $\mathbf{r}_1^{(t-1)},\mathbf{r}_2^{(t-1)} \in \mathbb{R}^{M_{\rm S} \times 1}$ are element-wise uniform random vectors sampled from $\mathcal{U}(0,1)$ to promote swarm diversity, $\omega$ is the inertia coefficient, and $c_1,c_2$ are the personal and global learning coefficients, respectively~\cite{piMultiuserCommunicationsMovableAntenna2023a}.
Finally, with the updated speed $\mathbf{s}_i^{(t)}$, the position of the $i$-th particle is updated as
\begin{equation}
    \label{eq:position_update}
    \mathbf{q}_i^{(t)} = \mathbf{q}_i^{(t-1)} + \mathbf{s}_i^{(t)}, \forall i \in \mathcal{I}_{\rm P}.
\end{equation}
By updating candidate subarray-positions in each PSO iteration following the procedures in (16)--\eqref{eq:position_update}, a suboptimal solution to problem (P3), denoted by $\mathbf{q}^*$, is obtained after $I_{\rm T}$ PSO iterations.

\subsubsection{Optimization of Antenna-Positions} 
\label{sec:opt_antenna}
For ease of exposition, we define $\mathbf{d}\triangleq [\mathbf{d}_1^T,\mathbf{d}_2^T,...,\mathbf{d}_{M_{\rm S}}^T]^T\in \mathbb{R}^{M\times 1}$ accounting for all per-subarray MA-positions $\{\mathbf{d}_s\}$. Given any optimized subarray-positions $\mathbf{q}^*$, problem (P2) reduces to the following problem for optimizing MA-positions $\mathbf{d}$
\begin{subequations}
    \begin{align}
 \label{eq:problem_array_antenna}
 (\text{P4})\quad &\max_{\mathbf{d}} \quad \check{R}(\mathbf{q}^*,\mathbf{d}) \\
 & \;\text{s.t.}  \quad \;\; \text{\eqref{eq:cons_intra_antenna}, \eqref{eq:cons_intrq}} ,\notag 
    \end{align}
\end{subequations}
where $\check{R}(\mathbf{q}^*,\mathbf{d}) \triangleq \tilde{R}(\mathbf{q}^*,\{\mathbf{d}_s\})$.
Similar to problem (P3), this problem is also challenging to solve optimally due to its non-convexity.
To tackle this difficulty, we also employ the PSO algorithm to find a high-quality solution to problem~(P4); the framework of which is similar to that in Section \ref{sec:opt_sub_array}.
Specifically, at the beginning of the antenna-wise PSO algorithm, a swarm of $\check{I}_{\rm P}$ (whose index set is denoted by $\check{\mathcal{I}}_{\rm P}\triangleq \{1,\dots, \check{I}_{\rm P}\}$) feasible position vectors are generated, each denoted by $\mathbf{d}_i^{(0)}\in \mathbb{R}^{M \times 1}, i \in \check{\mathcal{I}}_{\rm P}$, with $\check{\mathbf{s}}_i^{(0)}\in \mathbb{R}^{M \times 1}, i \in \check{\mathcal{I}}_{\rm P}$ denoting its initial speed. Then, a customized fitness function $\mathcal{F}_{\rm A}(\mathbf{d})$ is designed as follows to evaluate the quality of an arbitrary MA-position $\mathbf{d}$
\begin{equation}
    \label{eq:fitness_function_antenna}
    \mathcal{F}_{\rm A}(\mathbf{d}) = \tilde{R}(\mathbf{q}^*,\mathbf{d}) - \kappa P_{\rm A}(\mathbf{d}),
\end{equation}
where the first term corresponds to the objective function of problem (P4) and $\kappa P_{\rm A}(\mathbf{d})$ is the antenna-wise penalty function introduced to account for constraints \eqref{eq:cons_intra_antenna} and \eqref{eq:cons_intrq}. Specifically, $P_{\rm A}(\mathbf{d})$ is given by
\begin{align}
    P_{\rm A}(\mathbf{d}) &= \sum_{s=1}^{M
_{\rm S}} \sum_{a=1}^{M_{\rm A}}\left(\left[d_{s,a}-\left(L_{\rm A}-\frac{\lambda}{4}\right)\right]^+ 
    \!\!\!\!+\left[\frac{\lambda}{4} - d_{s,a}\right]^+ \right)\notag\\
    &+ \sum_{s=1}^{M_{\rm S}} \sum_{a=1}^{M_{\rm A}-1}\left[\frac{\lambda}{2}-\left(d_{s,a+1}-d_{s,a}\right)\right]^+.\label{eq:penalty_function_antenna}
\end{align}
Note that the first term in \eqref{eq:penalty_function_antenna} penalizes the constraint \eqref{eq:cons_intra_antenna} and the second term accounts for half-wavelength antenna-spacing constraint \eqref{eq:cons_intrq}.

Based on the designed fitness function $\mathcal{F}_{\rm A}$ in (21), the per-subarray MA-positions $\mathbf{d}_i^{(t)}$ can be updated in a similar manner to that in \eqref{eq:position_update} (see Section~\ref{sec:opt_sub_array}). Let $\mathbf{d}^*$ denote the suboptimal solution to problem (P4), which is obtained after $\check{I}_{\rm T}$ iterations. By alternately optimizing $\mathbf{q}$ and $\{\mathbf{d}_s\}$ (equivalently $\mathbf{d}$) until convergence, a suboptimal solution to problem (P2) can be obtained.

\begin{table}[t]
\centering
\caption{Simulation parameters}
\begin{tabular}{@{}ll|ll|ll@{}}
\toprule
\textbf{Parameter} & \textbf{Value} & \textbf{Parameter} & \textbf{Value} & \textbf{Parameter} & \textbf{Value} \\
\midrule
$f_c$ & 10{ GHz} & $\frac{P_{\rm t} \varpi^2 }{\sigma^2}$ & 9.78{ dB} & $L$ & $24\lambda$ \\
$M_{\rm S}$ & 4 & $L_{\rm A}$ & $\alpha\frac{L}{ M_{\rm S}}$ & $\alpha$ & $\frac{3}{8}$ \\
$\kappa$ & $10^6$ & $c_1, c_2$ & 2 & $\omega$ & 0.9 \\
$I_{\rm P}=\check{I}_{\rm P}$ & 300 & $I_{\rm T}=\check{I}_{\rm T}$ & 200 & $N_{\rm PA}$ & 3\\
\bottomrule
\end{tabular}
\label{tab:sim_param}
\vspace{-12pt}
\end{table}
\begin{remark}[Algorithm convergence and computational complexity]
    \label{sec:convergence}
    Since the objective function of problem (P2) is non-decreasing over each PSO iteration (and thus over each AO iteration), the convergence of proposed AO-based PSO algorithm is guaranteed~\cite{piMultiuserCommunicationsMovableAntenna2023a}. Next, the computational complexities of the proposed PSO algorithm for optimizing subarray-positions $\mathbf{q}$ and MA-positions $\{\mathbf{d}_s\}$ are in the orders of $\mathcal{O}(I_{\rm P} I_{\rm T})$ and $\mathcal{O}(\check{I}_{\rm P} \check{I}_{\rm T})$, respectively. In addition, the complexity of MMSE receive beamforming is $\mathcal{O}(M^3)$. As such, the overall complexity of the proposed AO-based PSO algorithm is $\mathcal{O}((I_{\rm P} I_{\rm T}+\check{I}_{\rm P} \check{I}_{\rm T}) M^3 I_{\rm A})$, where $I_{\rm A}$ is the number of AO iterations.
\end{remark}
\vspace{-1.2em}
\section{Numerical Results}
\begin{figure*}[h]
    \centering
    \begin{minipage}[b]{0.325\textwidth}
        \centering
        \includegraphics[width=\linewidth]{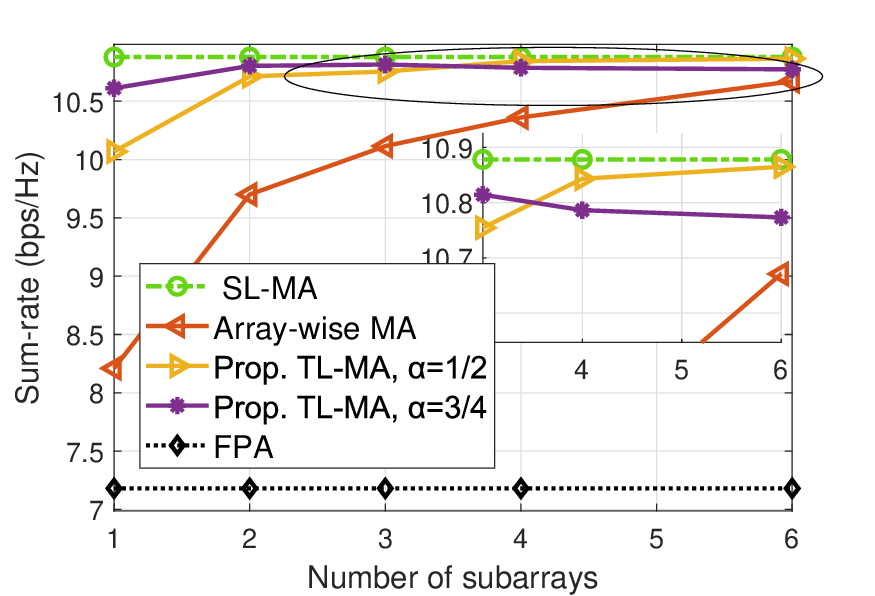}
        \caption{Sum-rate vs. number of subarrays $M_{\rm S}$.}
        \label{fig:rate_Msub}
    \end{minipage}\hfill
    \begin{minipage}[b]{0.325\textwidth}
        \centering
        \includegraphics[width=\linewidth]{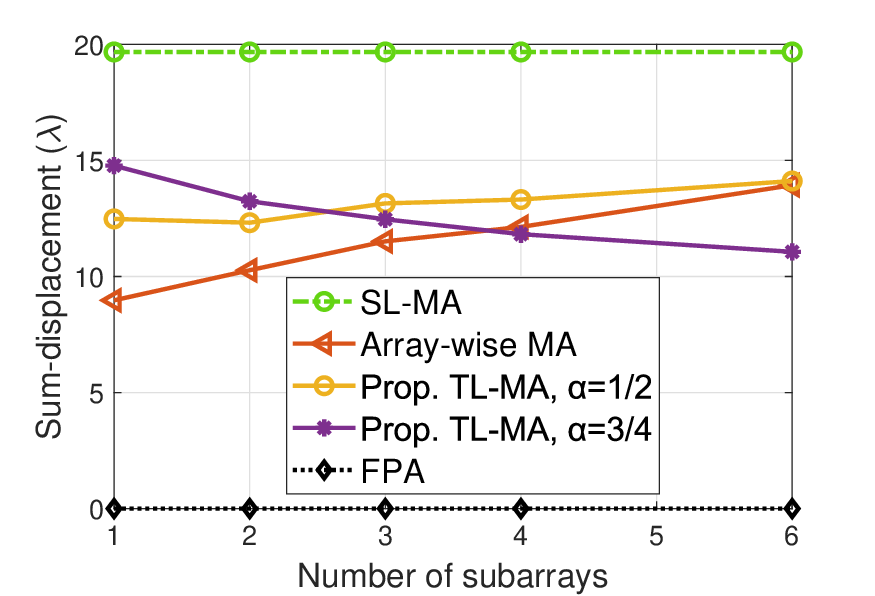}
        \caption{Sum-displacement vs. number of subarrays $M_{\rm S}$.}
        \label{fig:dist_Msub}
    \end{minipage}\hfill
    \begin{minipage}[b]{0.325\textwidth}
        \centering
        \includegraphics[width=\linewidth]{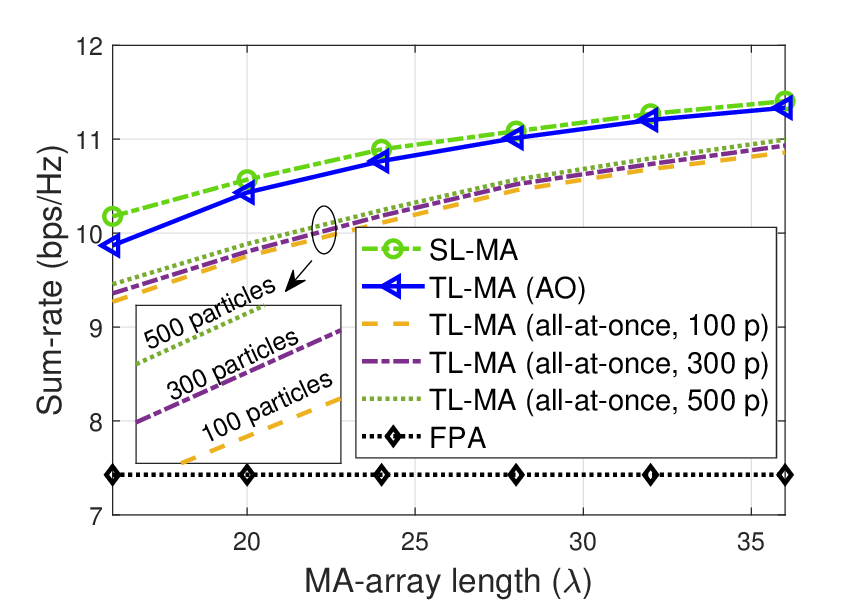}
        \caption{Sum-rate vs. MA-array length $L$ (100 p: $I_{\rm P}=100$).}
        \label{fig:rate_power}
    \end{minipage}
    \vspace{-20pt}
    \end{figure*}
\label{sec:numeric_results}

In this section, we evaluate the performance of our proposed TL-MA architecture and AO-based PSO algorithm. 
Specifically, a TL-MA based BS equipped with $M = 12$ antennas serves $K =3$ users. We assume that the spatial angles of all users follow $\theta_{k,n} \sim \mathcal{U}(-0.5,0.5)$ rad and their path gains are set as $\beta_{k,n}\sim \mathcal{CN}(0,\frac{\varpi^2}{N_{\rm PA}})$, with $\varpi$ being the average power~\cite{totarial}.
Moreover, the array length of each subarray is set as $L_{\rm A}=\frac{\alpha L}{M_{\rm S}}$, where $\alpha\in(\frac{M \lambda}{2 L},1)$ is a parameter controlling the allowable region of per-subarray movement. It is worth noting that the array-wise MA architecture in~\cite{gma} is a \emph{special case} of the proposed TL-MA architecture with $\alpha=\frac{M \lambda}{2 L}$ (i.e., $1/4$ in this case).
Unless otherwise specified, other parameters are set as Table~\ref{tab:sim_param}. 
For performance comparison, the following benchmark schemes are considered: 1) the element-wise SL-MA scheme in~\cite{piMultiuserCommunicationsMovableAntenna2023a}; 2) the array-wise scheme in~\cite{gma} (i.e., TL-MA w/ $\alpha=\frac{1}{4}$); and
3) the conventional FPA scheme.

In Fig.~\ref{fig:rate_Msub}, we show the system sum-rate versus the number of subarrays $M_{\rm S}$. 
Several key observations are made as follows.
First, all the MA-based schemes achieve significant sum-rate improvement over the conventional FPA scheme, thanks to the additional DoFs introduced by position adjustment. 
Second, compared with the array-wise scheme, the proposed TL-MA scheme with different subarray lengths achieves better sum-rate performance. This is because the array-wise scheme adjusts large-scale movement only, while the TL-MA scheme can further fine-tune the positions of MAs, thereby offering more spatial DoFs. 
Third, increasing the number of subarrays (i.e., $M_{\rm S}$) can improve the sum-rate performance when $M_{\rm S}$ is small. However, it may incur a slight rate loss when $M_{\rm S}$ is sufficiently large (i.e., $M_{\rm S}\ge 3$ for $\alpha=3/4$ in~Fig.~\ref{fig:rate_Msub}). 
This is expected since for the latter case, a large number of subarrays reduces the spatial DoFs available for small-scale fine-tuning (especially when $\alpha$ is large), although it can increase the DoFs in large-scale movement. This phenomenon reveals an inherent trade-off between large-scale and small-scale movements in the proposed TL-MA scheme.

In addition, we present the sum-displacement versus the number of subarrays $M_{\rm S}$ in Fig.~\ref{fig:dist_Msub}.
First, it is observed that although the SL-MA scheme achieves the highest sum-rate, it also incurs a significantly larger sum-displacement. 
In contrast, the proposed TL-MA architecture efficiently reduces the sum-displacement, 
while maintaining comparable rate performance with the SL-MA.
Second, the TL-MA with different values of $\alpha$ exhibits different sum-displacement trends as $M_{\rm S}$ increases. 
Specifically, a larger $\alpha$ (e.g., $\alpha=\frac{3}{4}$) yields a larger fine-tuning movable region for MAs, making it the dominant contributor to the sum-displacement. The fine-tuning region (i.e., the length of each subarray) decreases as $M_{\rm S}$ increases, thus leading to a decreasing sum-displacement trend.
In contrast, when $\alpha$ is small (e.g., $\alpha=\frac{1}{2}$ or $\frac{1}{4}$), the movable region for fine-tuning is limited, making the large-scale movement the main contributor.
In this regime, the increasing trend of sum-displacement is expected since more subarrays introduce more large-scale movement.
As such, for small subarray length architecture, decreasing $M_{\rm S}$ lowers sum-displacement but degrades rate; while for large subarray length architecture, increasing $M_{\rm S}$ is helpful for reducing sum-displacement but may reduce rate when $M_{\rm S}$ is large (see Fig.~\ref{fig:rate_Msub} for rate performance).

Last, in Fig.~\ref{fig:rate_power}, we compare the sum-rates of different schemes versus MA-array length $L$.
When the MA-array length increases from $16\lambda$ to $24\lambda$, all MA-based schemes achieve increased sum-rates because the enlarged movable region enhances the spatial DoFs.
Second, the proposed AO-based PSO algorithm achieves superior rate performance over the all-at-once PSO algorithm, even when the latter is associated more PSO particles.
This is because in the all-at-once scheme, all variables (i.e., subarray-positions and MA-positions) are simultaneously optimized, which not only enlarges the dimensionality of variables, but also introduces complicated position constraints (see constraints (2)--(3)) that are coupled with variables, making the particle exploration more difficult.
In contrast, the proposed AO-based PSO algorithm alternately optimizes the subarray-positions and MA-positions, reducing the dimensionality of each subproblem and decoupling the constraints from variables, thus facilitating the particle exploration.

\vspace{-1em}
\section{Conclusions}
\vspace{-0.2em}
In this letter, we proposed a new TL-MA architecture where the movement of each MA is decomposed into the large-scale movement of subarrays and small-scale movement of per-subarray MAs.
We formulated an optimization problem to maximize the uplink sum-rate. To solve this highly non-convex problem, we proposed an AO-based PSO algorithm, where the subarray positions and MA positions are optimized iteratively, given the optimal receive beamforming.
Numerical results validated that the proposed TL-MA architecture significantly reduces the sum-displacement of SL-MA, thus alleviating the control complexity and hardware cost, while achieving comparable rate performance. 
\vspace{-1.2em}
\bibliographystyle{IEEEtran}
\bibliography{bib_lib.bib}

\begin{thebibliography}{10}
\providecommand{\url}[1]{#1}
\csname url@samestyle\endcsname
\providecommand{\newblock}{\relax}
\providecommand{\bibinfo}[2]{#2}
\providecommand{\BIBentrySTDinterwordspacing}{\spaceskip=0pt\relax}
\providecommand{\BIBentryALTinterwordstretchfactor}{4}
\providecommand{\BIBentryALTinterwordspacing}{\spaceskip=\fontdimen2\font plus
\BIBentryALTinterwordstretchfactor\fontdimen3\font minus \fontdimen4\font\relax}
\providecommand{\BIBforeignlanguage}[2]{{%
\expandafter\ifx\csname l@#1\endcsname\relax
\typeout{** WARNING: IEEEtran.bst: No hyphenation pattern has been}%
\typeout{** loaded for the language `#1'. Using the pattern for}%
\typeout{** the default language instead.}%
\else
\language=\csname l@#1\endcsname
\fi
#2}}
\providecommand{\BIBdecl}{\relax}
\BIBdecl

\bibitem{you2024generationadvancedtransceivertechnologies}
C.~You, Y.~Cai, Y.~Liu, M.~Di~Renzo, T.~M. Duman, A.~Yener, and A.~Lee~Swindlehurst, ``Next generation advanced transceiver technologies for {6G} and beyond,'' \emph{IEEE J. Sel. Areas Commun.}, vol.~43, no.~3, pp. 582--627, Mar. 2025.

\bibitem{FAStotorial}
K.-K. Wong, A.~Shojaeifard, K.-F. Tong, and Y.~Zhang, ``Fluid antenna systems,'' \emph{IEEE Trans. Wireless Commun.}, vol.~20, no.~3, pp. 1950--1962, Nov. 2021.

\bibitem{graph}
W.~Mei, X.~Wei, B.~Ning, Z.~Chen, and R.~Zhang, ``Movable-antenna position optimization: A graph-based approach,'' \emph{IEEE Wireless Commun. Lett.}, vol.~13, no.~7, pp. 1853--1857, Apr. 2024.

\bibitem{RA}
X.~Shao, W.~Mei, C.~You, Q.~Wu, B.~Zheng, C.-X. Wang, J.~Li, R.~Zhang, R.~Schober, L.~Zhu, W.~Zhuang, and X.~Shen, ``A tutorial on six-dimensional movable antenna for {6G} networks: Synergizing positionable and rotatable antennas,'' \emph{IEEE Commun. Surv. Tutor.}, 2025, {Early Access}.

\bibitem{rotatableAntennaModelingOptimization2025}
B.~Zheng, Q.~Wu, T.~Ma, and R.~Zhang, ``Rotatable antenna enabled wireless communication: Modeling and optimization,'' \emph{arXiv preprint arXiv:2501.02595}, 2025.

\bibitem{pinchingA}
Z.~Ding, R.~Schober, and H.~Vincent~Poor, ``Flexible-antenna systems: A pinching-antenna perspective,'' \emph{IEEE Trans. Commun.}, 2025, {Early Access}.

\bibitem{totarial}
L.~Zhu, W.~Ma, W.~Mei, Y.~Zeng, Q.~Wu, B.~Ning, Z.~Xiao, X.~Shao, J.~Zhang, and R.~Zhang, ``A tutorial on movable antennas for wireless networks,'' \emph{IEEE Commun. Surv. Tutor.}, 2025, {Early Access}.

\bibitem{maMIMOCapacityCharacterization2024}
W.~Ma, L.~Zhu, and R.~Zhang, ``{MIMO} capacity characterization for movable antenna systems,'' \emph{IEEE Trans. Wireless Commun.}, vol.~23, no.~4, pp. 3392--3407, Sep. 2024.

\bibitem{piMultiuserCommunicationsMovableAntenna2023a}
Z.~Xiao, X.~Pi, L.~Zhu, X.-G. Xia, and R.~Zhang, ``Multiuser communications with movable-antenna base station: Joint antenna positioning, receive combining, and power control,'' \emph{IEEE Trans. Wireless Commun.}, vol.~23, no.~12, pp. 19\,744--19\,759, Nov. 2024.

\bibitem{MAimplementation}
B.~Ning, S.~Yang, Y.~Wu, P.~Wang, W.~Mei, C.~Yuen, and E.~Bjornson, ``Movable antenna-enhanced wireless communications: General architectures and implementation methods,'' \emph{IEEE Wireless Commun.}, 2025, {Early Access}.

\bibitem{gma}
H.~Lu, Y.~Zeng, S.~Jin, and R.~Zhang, ``Group movable antenna with flexible sparsity: Joint array position and sparsity optimization,'' \emph{IEEE Wireless Commun. Lett.}, vol.~13, no.~12, pp. 3573--3577, Oct. 2024.

\bibitem{robust}
B.~Lyu, C.~Zhou, S.~Gong, D.~T. Hoang, and Y.-C. Liang, ``Robust secure transmission for active {RIS} enabled symbiotic radio multicast communications,'' \emph{IEEE Trans. Wireless Commun.}, vol.~22, no.~12, pp. 8766--8780, Apr. 2023.

\end{thebibliography}

\end{document}